\documentclass[preprint,review,12pt]{elsarticle}

\usepackage{amssymb}

\usepackage{amsmath,amssymb,amsfonts}
\usepackage{algorithmic}
\usepackage{graphicx}
\usepackage{ulem}
\usepackage{soul}
\usepackage{textcomp}
\usepackage{xcolor}
\usepackage{float}
\usepackage{overpic}
\usepackage{multirow}
\usepackage{subfigure}
\usepackage{booktabs}
\usepackage{bbding}

\usepackage[colorlinks,
linkcolor=red,
anchorcolor=red,
citecolor=green]{hyperref}
\def\BibTeX{{\rm B\kern-.05em{\sc i\kern-.025em b}\kern-.08em
		T\kern-.1667em\lower.7ex\hbox{E}\kern-.125emX}}

\journal{Neural Networks}

\begin{document}

\title{Multi-modal Cross-domain Self-supervised Pre-training for fMRI and EEG Fusion}

\author[1]{Xinxu Wei}
\ead{xiw523@lehigh.edu}

\author[2]{Kanhao Zhao}
\ead{kaz220@lehigh.edu}

\author[2]{Yong Jiao}
\ead{yoj323@lehigh.edu}

\author[3]{Nancy B. Carlisle}
\ead{nbc415@lehigh.edu}

\author[4]{Hua Xie}
\ead{HXIE@childrensnational.org}

\author[5]{Gregory A. Fonzo}
\ead{gfonzo@austin.utexas.edu}

\author[2,1]{Yu Zhang*}
\ead{yuzi20@lehigh.edu}


\cortext[cor1]{Corresponding author}

\address[1]{Department of Electrical and Computer Engineering, Lehigh University, Bethlehem, PA 18015, USA}

\address[2]{Department of Bioengineering, Lehigh University, Bethlehem, PA 18015, USA}

\address[3]{Department of Psychology, Lehigh University, Bethlehem, PA 18015, USA}

\address[4]{Center for Neuroscience Research, Children's National Hospital, Washington, DC 20010, USA}

\address[5]{Center for Psychedelic Research and Therapy, Department of Psychiatry and Behavioral Sciences, Dell Medical School, The University of Texas at Austin, Austin, TX 78712, USA}

\begin{abstract}
Neuroimaging techniques including functional magnetic resonance imaging (fMRI) and electroencephalogram (EEG) have shown promise in detecting functional abnormalities in various brain disorders. However, existing studies often focus on a single domain or modality, neglecting the valuable complementary information offered by multiple domains from both fMRI and EEG, which is crucial for a comprehensive representation of disorder pathology. This limitation poses a challenge in effectively leveraging the synergistic information derived from these modalities. To address this, we propose a Multi-modal Cross-domain Self-supervised Pre-training Model (MCSP), a novel approach that leverages self-supervised learning to synergize multi-modal information across spatial, temporal, and spectral domains. Our model employs cross-domain self-supervised loss that bridges domain differences by implementing domain-specific data augmentation and contrastive loss, enhancing feature discrimination. Furthermore, MCSP introduces cross-modal self-supervised loss to capitalize on the complementary information of fMRI and EEG, facilitating knowledge distillation within domains and maximizing cross-modal feature convergence. We constructed a large-scale pre-training dataset and pretrained MCSP model by leveraging proposed self-supervised paradigms to fully harness multimodal neuroimaging data. Through comprehensive experiments, we have demonstrated the superior performance and generalizability of our model on multiple classification tasks. Our study contributes a significant advancement in the fusion of fMRI and EEG, marking a novel integration of cross-domain features, which enriches the existing landscape of neuroimaging research, particularly within the context of mental disorder studies.

\end{abstract}

\begin{keyword}
Multi-modal fusion, Self-supervised learning, Pre-training, Knowledge transfer, Neuroimaging
\end{keyword}

\maketitle

\section{Introduction}
\label{sec:introduction}
Functional magnetic resonance imaging (fMRI) and Electroencephalography (EEG) are widely-used neuroimaging techniques for investigating brain disorders. 
Both of these two modalities offer insights into brain function across spatial, temporal, and spectral domains.
In each modality, the three domains possess distinct characteristics. For example, fMRI measures low-frequency fluctuations in blood oxygen level-dependent (BOLD) signals\cite{yin2022deep} with high spatial resolution allowing for the detailed detection of neural activity across brain regions. In contrast, EEG, which records electric currents generated by ensembles of postsynaptic potentials induced from corresponding neurons' activities with high temporal resolution, is more sensitive to detecting instantaneous fluctuations of cerebral electrical activity \cite{lawhern2018eegnet}. fMRI's high spatial resolution is advantageous for characterizing the brain as spatial functional graph networks \cite{gardumi2016effect}. On the other hand, EEG's superior temporal resolution captures brain electrical activity at the millisecond level, making it well-suited for studying rapid dynamic changes in brain activity \cite{burle2015spatial}.

A growing body of work \cite{zhang2022self, yang2023mapping} has demonstrated that integrating multiple domains and modalities may be crucial for analyzing the neuroimaging data. Despite increasing research efforts on integrating multi-domain and multi-modal features, many of them simply concatenated features from these domains or modalities, ignoring the intrinsic complementary properties between them\cite{yu2016building}. This perspective fails to consider the possible interactions between domains of a single modality and the valuable complementary information that different modalities within each domain could offer. In contrast, the fusion of features from different domains can provide the model with opportunities to learn diverse representations of the same input data. The integration of fMRI and EEG data, despite their differing measurement principles and resolutions, can provide a multi-scale and synergistic understanding of brain activity. So far, there have been limited deep learning designs for elegantly integrating these two modalities. We argue that simply concatenating features from different domains within a single modality or concatenating features from fMRI and EEG modalities alone does not yield satisfactory results for neuroimaging classification tasks. 

To address this challenge, we propose a novel algorithm, named Multi-modal Cross-domain Self-supervised Pre-training Model (MCSP). To fully capture latent interactions among domains within each modality rather than simply concatenating features from different domains, we introduce the Cross-domain Self-Supervised Loss function (CD-SSL). This function utilizes domain-specific augmentation and contrastive learning techniques to maximize similarity among domains.
Furthermore, for multi-modal fusion, instead of directly concatenating features from different modalities, we propose the Cross-modal Self-Supervised Loss function (CM-SSL). CM-SSL explicitly leverages the complementary characteristics between fMRI and EEG within each domain by distilling richer information to the other.
To demonstrate the wide applicability and flexibility of our model, we conducted cross-model distillation across domains and analyzed potential universal pre-training abilities.

The major contributions of our study are as follows:

\begin{itemize}
	
	\item We develop the innovative Multi-modal Cross-domain Self-supervised Pre-training Model (MCSP) to harness and integrate the unique spatial, temporal, and frequency domains of multi-modal neuroimaging data, specifically fMRI and EEG.
	
	\item We design a cross-domain self-supervised loss function (CD-SSL) for the pre-trained model. This function employs contrastive learning and domain-specific data augmentation to optimize the exploration of multi-domain data, enhancing the cross-domain similarity of features.
	
	\item We introduce a cross-modality self-supervised loss function (CM-SSL) that efficiently mines data across modalities, focusing on maximizing the similarity between them. Additionally, we utilize cross-modal distillation techniques to capitalize on the complementary aspects of fMRI and EEG within each domain.

 	\item To validate the efficacy of our model, we have conducted comprehensive experiments on multiple neuroimaging datasets. The experimental results underscore the superior performance and potential of our approach.
	
\end{itemize}

Our methodology demonstrates the potential to advance multimodal neuroimaging analysis by creating a cohesive framework that maximizes the utility of both spatially rich and temporally precise neuroimaging data, setting a new standard for multi-modal neuroimaging research.

\section{Related Works}
\label{sec:related works}


\subsection{Deep Learning for Brain Networks Learning}

Various methods have been developed for neuroimaging analysis \cite{zhang2021survey}, encompassing both fMRI \cite{cui2022braingb}\cite{kan2022fbnetgen} and EEG \cite{craik2019deep}. A noteworthy trend is the recent surge in deep learning and graph convolutional networks (GCNs) \cite{kipf2016semi}. Regarding fMRI signals, BrainNetCNN \cite{kawahara2017brainnetcnn} pioneered the use of Convolutional Neural Networks (CNNs) \cite{he2016deep} in brain network analysis, employing CNNs to extract deep semantic features from fMRI. BrainGNN \cite{li2021braingnn} explored brain network information by modeling the brain as a graph with nodes and edges, leveraging GCN to extract spatial features of the brain networks. Brain Network Transformer (BNT) \cite{kan2022brain} and Dynamic BNT \cite{kan2023dynamic} improved Graph Transformer \cite{yun2019graph} and applied it to extract fMRI features. Concerning EEG signals, EEGNet \cite{lawhern2018eegnet} introduced depthwise and separable convolutions to fully extract spatial and temporal features, achieving improved neural decoding performance. EEG-GNN \cite{demir2021eeg} adopted a graph network to learn geometric features by modeling electrodes as nodes in the graph.

While the aforementioned methods focus on extracting spatial or temporal information from fMRI or EEG signals separately, they often overlook the correlation between spatial and temporal information, as well as the complementary properties between fMRI and EEG. Our method bridges this gap by considering spatial, temporal, and frequency information across both modalities comprehensively.

\subsection{Multi-modal Fusion for Neuroimaging Integration}

In neuroscience, the existence of various neuroimaging modalities, such as fMRI and EEG, underlines the complexity and depth of brain study \cite{10182318,pustina2015predicting}. These modalities, each offering unique insights into brain activity and/or structure, are crucial for advancing our understanding and treatment of brain disorders. The integration of multi-modal data is pivotal for improving diagnosis and prognosis of brain disorders. Increasing research efforts have been dedicated to achieving this goal. For instance, a CNN-LSTM framework \cite{rabbani2023deep} was proposed to conduct the fusion of large volume time-series data like EEG. \cite{zhou2022interpretable} improved the performance of mild cognitive impairment diagnosis by integrating both MRI and PET modalities via GCN \cite{kipf2016semi}. Cross-GNN \cite{yang2023mapping} integrated fMRI and DTI to capture inter-modal dependencies through dynamic graph and mutual learning. A Multi-Modal and Multi-Atlas Integrated Framework \cite{long2022multi} was proposed to fuse resting-state fMRI (rs-fMRI) and structural MRI (sMRI). Moreover, the potential for fusion between fMRI and EEG has been demonstrated, suggesting that combining these modalities can yield significant benefits \cite{calhoun2016multimodal}.
However, most previous methods for integrating fMRI and EEG have not explicitly utilized the complementary information between them. Our method bridges this gap by proposing a novel deep learning framework that constructs self-supervised loss across different domains from both fMRI and EEG modalities, explicitly exploring their complementary information via knowledge distillation and consistency characteristics via contrastive regularization.

\subsection{Self-supervised Pre-training and Multi-domain Fusion}

Self-supervised learning \cite{jaiswal2020survey} explores latent data relationships by constructing self-supervised loss functions, a widely embraced practice in various fields. This method facilitates pre-training or optimizing training through inherent data characteristics without labeled examples. 

Recent advancements in the field include a self-supervised pre-training model, trained on large-scale neuroimaging data \cite{thomas2022self}, showcasing its potential in brain science. Additionally, a pre-training pipeline utilizing self-supervised learning techniques was introduced, aiming to regularize consistency loss across different domains (spatial and temporal) for time series prediction, encompassing EEG and MEG signals \cite{zhang2022self}. Similarly, a framework leveraging principles of self-supervised learning was developed, focusing on contrastive regularization between time series signals like EEG \cite{eldele2023self}. Furthermore, another model delved into learning pretext information to enhance spatiotemporal representations for emotion recognition from ECG signals \cite{sarkar2020self}.

While some of these methods consider constructing self-supervised information across different domains \cite{zhang2022self}, none have explicitly explored both multi-modal and multi-domain brain signals for self-supervised pre-training. In order to bridge this gap, our approach not only constructs self-supervised loss across domains but also explicitly leverages complementary information between different modalities to establish a novel self-supervised paradigm.

\begin{figure*}[htbp]
	\centering
	\includegraphics[width=\linewidth]{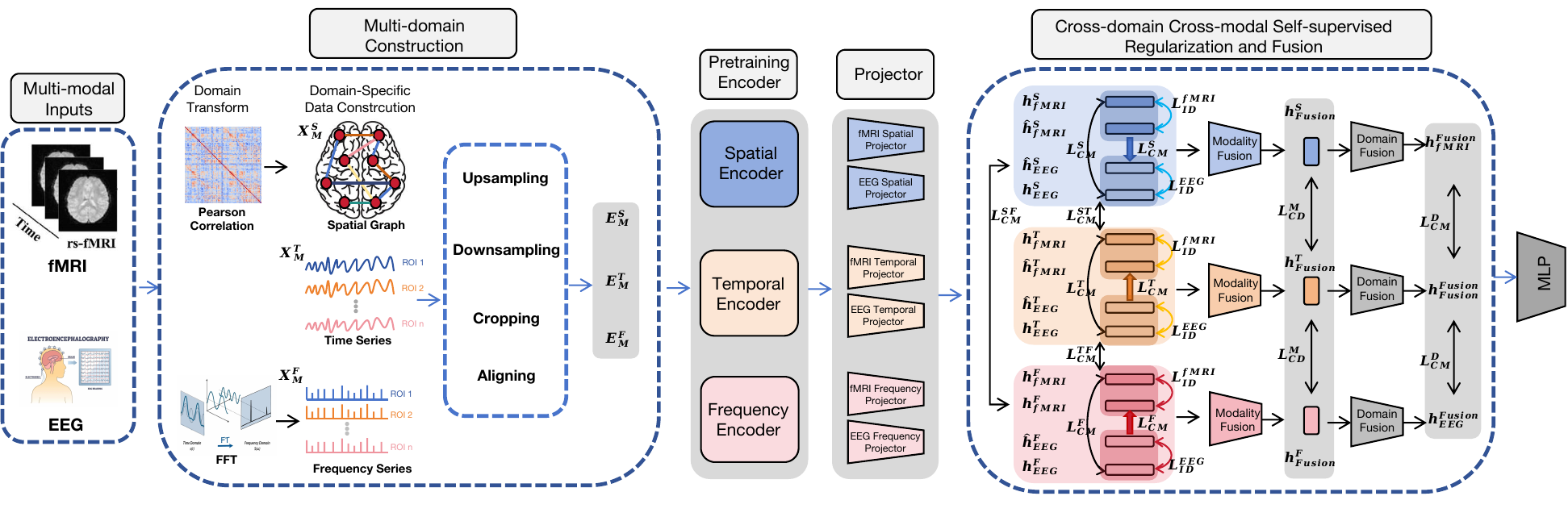}
	\caption{The network architecture of the proposed MCSP. The model consists of several components, including multimodal inputs, the construction of data from different domains, encoding and extraction of features from different domains, projection heads of different modalities data in different domains as well as cross-modal and cross-domain self-supervised constraints and fusion. And a task-agnostic MLP is adopted as the classifier.}
	\label{model}
\end{figure*}

\section{Methodology}
\label{sec:methods}

\subsection{Overall Architecture}

As illustrated in Fig. \ref{model}, our proposed MCSP model is designed to handle multi-modal and multi-domain inputs, specifically fMRI and EEG, for comprehensive feature extraction from brain signals across spatial, temporal, and frequency domains. 
During the pre-training phase, based on contrastive self-supervised pre-training fashion \cite{zhang2022self} \cite{jaiswal2020survey}, we introduce Cross-domain self-supervised loss (CD-SSL) and Cross-modal self-supervised loss (CM-SSL) to maximize the similarity of features from different domains and modalities. Additionally, we incorporate cross-modal knowledge distillation to explicitly leverage the complementary information between fMRI and EEG across different domains.

\subsection{Domain-specific Data Construction}
Before feeding the data into domain-specific encoders, we first preprocess and construct the data for each domain. In the spatial domain, after data preprocessing, we compute connectivity from both fMRI and EEG source estimates, each containing time series signals extracted from 100 regions of interest (ROIs). The data preprocessing and connectivity calculation methods are described in Section \ref{sec:datasets}. For fMRI, following \cite{li2021braingnn}, we construct functional connectome using the Pearson Correlation Coefficients as node features and adopt the correlation as edge connections. For EEG, we compute the power envelope connectivity and use power envelope coefficients as node features. Similarly, we adopt the correlation as edge connections.
For time series signals, we standardize the fMRI time series with varying lengths to the unified $E^{T}_{fMRI} \in \mathbb{R}^{N \times l}$. Here, $N$ represents the number of ROIs, and $l$ represents the unified length of sequences. Specifically, we set $N = 100$ and $l = 200$ for fMRI time series.
As for EEG data, given that EEG time series are typically much longer than fMRI time series due to a higher sampling rate and contain more precise information due to a higher resolution, we initially downsample and unify the EEG time series for each subject to $E^{T}_{EEG} \in \mathbb{R}^{N \times l'}$ with unified length, where $l' = 25000$. Subsequently, we divide the entire time series sequence into 125 segments $E^{T}_{EEG} \in \mathbb{R}^{N \times l}$, each having the same length of $l = 200$, to align with the time series of fMRI data. 
In addition, for frequency-domain sequences, we apply the same processing approach as for time-domain sequences. We perform upsampling and downsampling on the fMRI frequency-domain sequences to obtain a uniformly sized $E^{F}_{fMRI} \in \mathbb{R}^{N \times l}$. Similarly, we align and segment the EEG frequency-domain sequences to obtain $E^{F}_{EEG} \in \mathbb{R}^{N \times l}$. Note that for each subject, both $E^{T}_{EEG}$ and $E^{F}_{EEG}$ contain a set of sub-sequences.

\subsection{Domain-specific Encoders and Projectors}
\subsubsection{Modality-agnostic Domain-specific Encoders}

We utilize modality-agnostic domain-specific encoders to extract unique features for each domain. In the spatial domain, we employ a Graph Transformer model with positional embeddings \cite{yun2019graph} as the encoder to capture spatial information by treating each ROI as a node, thereby capturing the spatial distribution of fMRI and EEG, as well as connectivity features between ROIs. For the temporal and frequency domain, Transformer models are used to extract time series features while considering dependencies between long-distance time points. 
\begin{equation}
	\begin{aligned}
		&{e}^{D}_{M} = Encoder^{D}_{m}(E^{D}_{M}), \quad {E}^{D}_{M} \in \mathbb{R}^{K \times c}
	\end{aligned}
	\label{encoder}
\end{equation}
where $M$ means modality and $M\in \{MRI, EEG\}$. $D$ represents domain and $D\in \{Spatial, Temporal, Frequency\}$. $m$ means type of encoder model and $m\in \{Graph Transformer, Transformer, Transformer\}$ for three domains, respectively. For spatial domain, $K$ refers to the number of nodes, while $c$ represents the dimension of the node feature. For temporal and frequency domains, $K$ means the number of brain ROIs, while $c$ denotes the length of time and frequency sequences.
After extracting features via domain-specific encoders, we can obtain six kinds of hidden-layer feature embeddings ${e}^{D}_{M}$ from three domains and two modalities. All these extracted feature embeddings will be fed into projectors. Note that for each subject, both ${e}^{T}_{EEG}$ and ${e}^{F}_{EEG}$ represent a set of feature embeddings. 

\subsubsection{Modality-aware Projection Heads}
\begin{figure}[htbp]
	\centering
	\includegraphics[width=12cm]{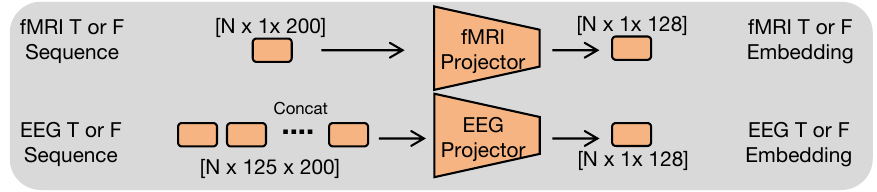}
	\caption{The temporal \& frequency projectors for aligning fMRI and EEG sequences in the embedding space. For EEG's time and frequency sequences, we concatenate all the segments of features outputted by the encoder and then input them into the projector. $N$ denotes the number of subjects. $'125'$ represents the 125 equally sized segments we can obtain for each subject's EEG time and frequency sequences. Based on the experimental results, setting the length to 200 strikes an optimal balance between performance and computational efficiency.}
	\label{pic_project}
\end{figure}

Projectors are employed to obtain contrastive embeddings, which are crucial for calculating the contrastive loss function and conducting classification via fully-connected layers. Due to the lengthy nature of the time and frequency sequences in EEG, in order to align them with fMRI, we segment long sequences from each subject into shorter, equal-length segments (length = 200) during the data construction process. As shown in Fig. \ref{pic_project}, for these segments, we concatenate them before inputting them into the projector. Then, we project them into a high-dimensional space with unified dimensions to obtain consistent embeddings ${h}^{D}_{M}$.
\begin{equation}
	\begin{aligned}
		&{h}^{D}_{M} = Projector^{D}_{m}(e^{D}_{M}), \quad {h}^{D}_{M} \in \mathbb{R}^{N \times 128}
	\end{aligned}
	\label{projector}
\end{equation}

\subsection{Cross-domain Self-supervised Loss}
Self-supervised learning enables the exploration of intrinsic relationships between features, uncovering deep dependencies within the data. The critical aspect of self-supervised learning lies in the construction of effective self-supervised loss functions and embeddings for loss computation.

To fully explore the latent interaction across different domains among a single modality, we propose a Cross-domain Self-supervised Loss, denoted as CD-SSL, which consists of an Intra-domain Cross-view Consistency Loss $L^{M}_{ID}$ and a Cross-domain Consistency Loss $L^{M}_{CD}$. 
\begin{figure}[htbp]
	\centering
	\includegraphics[width=\linewidth]{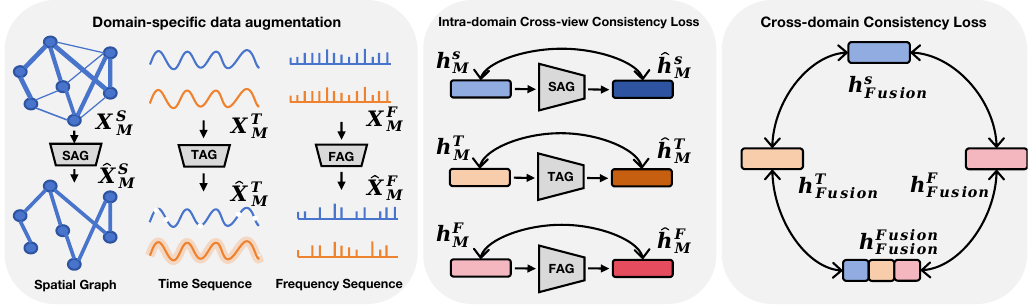}
	\caption{The proposed Cross-domain Self-supervised loss function, which consists of an Intra-domain Cross-view Consistency Loss $L^{M}_{ID}$ and a Cross-domain Consistency Loss $L^{M}_{CD}$.}
	\label{pic_cd_ssl}
\end{figure}
As shown in Fig. \ref{pic_cd_ssl}, for each domain, we utilize domain-specific augmentation techniques to generate pairwise contrastive embeddings. 
Inspired by graph augmentation \cite{zhao2021data} in graph contrastive learning (GCL) \cite{zhu2021graph}, for the spatial domain, we leverage graph augmentation to generate spatially augmented views while keep semantic invariance of the brain network graph. For structure-based graph augmentation, there are two options. First, we randomly remove 20$\%$ to 50$\%$ of the weakest edges' connection strength, resulting in graph views with the same node distribution and node features but sparser edges. Second, we apply a certain degree of random perturbation to relatively strong edges, ensuring that the perturbation added to edges does not alter the graph's semantics. This yields graph views with the same node distribution and edge sparsity, but with varying strengths for some edges. For the temporal domain, to obtain temporal augmented views, we introduce random noise to the time series of both fMRI and EEG. Additionally, considering the higher sampling rate of EEG's time series, we randomly discard time points in the EEG time series.
For the frequency domain, we directly transform the augmented temporal views from temporal domain to the frequency domain, and obtain the augmented frequency views.
The augmentation processing for the three domains is formulated as below:
\begin{equation}
	\begin{aligned}
		&{X}^{s}_{M}  \quad \xrightarrow{Remove Edge} \quad \widehat{X}^{s}_{M}, \quad
		{X}^{t}_{M}  & \xrightarrow[Add Noise]{Remove Point} \quad \widehat{X}^{t}_{M},
		 \quad \widehat{X}^{t}_{M}  \quad \xrightarrow{FFT} \quad \widehat{X}^{f}_{M} \\
	\end{aligned}
	\label{aug}
\end{equation}
After obtaining augmented views $\widehat h^{D}_{i}$ for each sample from each domain by feeding augmented samples $\widehat X^{D}_{M}$ into domain-specific encoders and projectors, we maximize the similarity between the original domain-specific anchor views $h^{D}_{i}$ and the augmented views $\widehat h^{D}_{i}$ in each domain $D$ while minimizing the similarity between the anchor samples $h^{D}_{i}$ and the negative samples $h^{D}_{j}$ by applying an intra-domain contrastive loss $L^{M}_{ID}(h^{D}_{i},\widehat h^{D}_{i})$ as below: 
\begin{equation}
	\mathcal L^{M}_{ID}(h^{D}_{i},\widehat h^{D}_{i}) =   -log\frac{exp(sim(z^{D}_{i}, \widehat z^{D}_{i})/\tau)}{\sum_{j=0}^{size-1}exp(sim(z^{D}_{i}, z^{D}_{j}))/\tau}
	\label{im_loss}
\end{equation}
where $z_i = softmax(h_i)$. $M$ means modality and $M\in \{MRI, EEG\}$. $D$ means doamin and $D\in\{Spatial, Temporal, Frequency\}$.

Meanwhile, as shown in Eq. \ref{cd_loss}, for different samples across different domains, we employ a cross-domain contrastive loss $L^{M}_{CD}(h^{D_{x}}_{i}, h^{D_{y}}_{i})$ to maximize the similarity between positive paired samples $\{h^{D_{x}}_{i}, h^{D_{y}}_{i}\}$ and minimize the similarity between negative paired samples $\{h^{D_{x}}_{i}, h^{D_{y}}_{j}\}$.
\begin{equation}
	\mathcal L^{M}_{CD}(h^{D_{x}}_{i}, h^{D_{y}}_{i}) =   -log\frac{exp(sim(z^{D_{x}}_{i}, z^{D_{y}}_{i})/\tau)}{\sum_{j=0}^{size-1}exp(sim(z^{D_{x}}_{i}, z^{D_{y}}_{j}))/\tau} 
	\label{cd_loss}
\end{equation}
where $z_i = softmax(h_i)$. $D_{x}$ and $D_{y}$ mean different Domains, that is $D_{x} \neq D_{y}$. $M$ means modality and $M\in \{MRI, EEG\}$.

Finally, we combine the two contrastive loss $L^{M}_{ID}(h^{D}_{i},\widehat h^{D}_{i})$ and $L^{M}_{CD}(h^{D_{x}}_{i}, h^{D_{y}}_{i})$ to form our $L^{M}_{CD-SSL}$ with a balance coefficient $\alpha=0.5$.
\begin{equation}
	\mathcal L^{M}_{CD-SSL} = \alpha \cdot \mathcal L^{M}_{ID} + (1 - \alpha) \cdot  \mathcal L^{M}_{CD}
	\label{cd_ssl}
\end{equation}

\subsection{Cross-modal Self-supervised Loss}

For capturing the inner interaction across different modalities during the pre-training or the online training, we propose a cross-modal self-supervised loss, named CM-SSL, which consists of an Intra-domain Cross-modal Distillation Loss $L^{D}_{IM}$ and a Cross-modal Consistency Loss $L^{D}_{CM}$.
\begin{figure}[htbp]
	\centering
	\includegraphics[width=11cm]{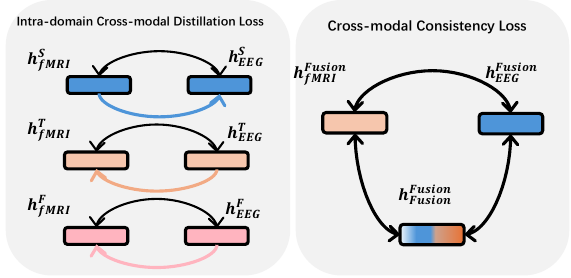}
	\caption{The proposed cross-modal self-supervised loss function, which consists of an Intra-domain Cross-modal Distillation Loss $L^{D}_{IM}$ and a Cross-modal Consistency Loss $L^{D}_{CM}$.}
	\label{cm_ssl}
\end{figure}
As illustrated in Fig. \ref{cm_ssl}, for the embeddings of fMRI and EEG from three domains, we rely on two assumptions. 
1). Assumption I, the similarity between embeddings of different modalities within the same domain should be as close as possible since they share similar semantics and can be treated as positive samples. 
2). Assumption II, because the information of fMRI and EEG signals varies across three domains, emphasizing different aspects (e.g., fMRI has higher spatial resolution but lower temporal and frequency resolution, while EEG has higher temporal and frequency resolution but lower spatial resolution), we aim to distill the knowledge from the modality with richer information into the other. Thus, we formulate the Intra-domain Cross-modal Distillation Loss $L^{D}_{IM}$ as below:
\begin{equation}
\footnotesize
	\begin{aligned}
	&\mathcal 
	L^{D}_{IM}(h^{M_{x}}_{i}, h^{M_{y}}_{i}) = -log\frac{exp(sim(z^{M_{x}}_{i}, z^{M_{y}}_{i})/\tau)}{\sum_{j=0}^{size-1}exp((sim(z^{M_{x}}_{i}, z^{M_{y}}_{j}) + sim(z^{M_{y}}_{i}, z^{M_{x}}_{j}))/\tau)} + KL(z^{M_{x}}_{i} ||  z^{M_{y}}_{i})
	\end{aligned}
	\label{id_loss}
\end{equation}
where $D$ means domain and $D\in \{Spatial, Temporal, Frequency\}$. $z_i = softmax(h_i)$.
$M_{x}$ and $M_{y}$ denote two different modalities. $x$ and $y$ $\in \{fMRI, EEG\}$ and $x \neq y$. $KL(.)$ stands for Kullback-Leibler divergence. $exp(.)$ represents the exponential function. $sim(.)$ denotes cosine similarity. $\tau$ indicates the temperature coefficient, and here we set it to 0.2. Note that when in the spatial domain (i.e., $D=Spatial$), $x=EEG$ and $y=fMRI$, this means that in the spatial domain, we expect knowledge from fMRI to distill into EEG, and EEG features should learn and approach those of fMRI. Conversely, when in the temporal or frequency domain (i.e., $D=Temporal$ or $Frequency$), $x=fMRI$ and $y=EEG$, we expect knowledge from EEG to distill into fMRI, and fMRI features should learn and approach those of EEG.

As illustrated in Fig. \ref{cm_ssl}, we also apply an overall distribution regularization across the domain-integrated fusion embedding of fMRI, EEG and their fused feature. 
\begin{equation}
	\begin{aligned}
	\mathcal L^{D}_{CM}&(h^{M_{x}}_{i}, h^{M_{y}}_{i})  =  KL(z^{M_{f}}_{i} || z^{M_{e}}_{i})  +
	KL(z^{M_{e}}_{i} || z^{M_{fe}}_{i}) +  KL(z^{M_{fe}}_{i} || z^{M_{f}}_{i}) 
	\end{aligned}
	\label{cd_loss_f}
\end{equation}
where $z_i = softmax(h_i)$.

Finally, we combine the Intra-modal regularization $L^{D}_{IM}(h^{M_{x}}_{i}, h^{M_{y}}_{i})$ and the Cross-modal regularization $L^{D}_{CM}(h^{M_{x}}_{i}, h^{M_{y}}_{i})$ to form our $L^{D}_{CM-SSL}$  with a balance coefficient $\alpha=0.8$.
\begin{equation}
	\begin{aligned}
	&\mathcal L^{D}_{CM-SSL} = \alpha \cdot \mathcal L^{D}_{IM} + (1 - \alpha) \cdot  \mathcal L^{D}_{CM}
	\end{aligned}
	\label{cm_ssl_loss}
\end{equation}

\subsection{Cross-model Distillation across Domains}

The proposed self-supervised loss functions are designed to fully explore the latent interaction among different modalities and domains, especially leveraging the complementary information between domains of fMRI and EEG modalities. 
During the re-training for the downstream classification tasks, the Cross Entropy (CE) loss $\mathcal L_{CE}$ is adopted as the loss function.

\begin{figure}[htbp]
	\centering
	\includegraphics[width=11cm]{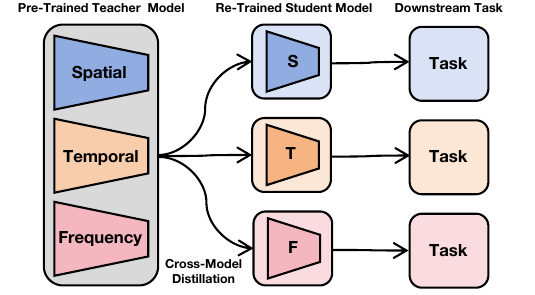}
	\caption{The diagram illustrates the concept of cross-model distillation pre-training across domains.}
	\label{distill}
\end{figure}

Additionally, as illustrated in Fig.\ref{distill}, another potential pre-training and fine-tuning paradigm is cross-model distillation across domains, showing the ability to distill the knowledge from a large teacher model, fused with knowledge from multiple domains, into a smaller student model focusing on a single domain.
The key lies in defining an appropriate loss function, which includes both soft target loss and hard target loss as shown below
\begin{equation}
	\mathcal{L}_{soft} = -\sum_{i}^{classes} p_i^{\text{STF}} \log(q_i^{\text{D}}), \quad \mathcal{L}_{hard} = -\sum_{i}^{classes} y_i \log(q_i^{\text{D}})
\end{equation}

where $STF$ means a large teacher model which is trained from the fusion of three domains. $D$ means the candidate student model from a single domain and $D \in \{S,T,F\}$.


\section{Data Acquisition and Preprocessing}
\subsection{Datasets}
\label{sec:datasets}
We evaluated the efficacy of the proposed model on multiple independent datasets, including ADHD-200 \cite{adhd2012adhd}, Autism Brain Imaging Data Exchange (ABIDE I \cite{di2014autism} and ABIDE II \cite{di2017enhancing}), Establishing Moderators and Biosignatures of Antidepressant Response in Clinical Care (EMBARC)\cite{trivedi2016establishing}, and Healthy Brain Network (HBN). 

\subsubsection{ADHD-200}
The ADHD-200 dataset\cite{adhd2012adhd} is a collection of rs-fMRI data for studying ADHD. It aggregates data from various research institutions, aiming to assist researchers in gaining deeper insights into the brain dysfunction of ADHD. In this study, we used rs-fMRI data, including 830 subjects and 1086 runs, which contain 404 ADHD patients and 682 healthy controls (HCs) runs after quality control, collected from eight sites.

\subsubsection{ABIDE}
The ABIDE \cite{di2014autism,di2017enhancing} is a collection of rs-fMRI data for studying ASD. This dataset comprises two phases: ABIDE I and ABIDE II. The ABIDE I  consists of rs-fMRI data from 555 ASD patient runs and 598 matched HC runs while the ABIDE II has collected data from 581 ASD patient runs and 733 HC runs. We evaluated our model for ASD prediction on the ABIDE I and II, respectively, to confirm its robustness against different study phases of the dataset.

\subsubsection{EMBARC}
The EMBARC \cite{trivedi2016establishing} is a multi-modal dataset containing both fMRI and EEG collected from 265 MDD patients. We utilized this dataset to examine multi-modal fusion and conduct two different prediction tasks: Depression Grading and Sex Classification. For Depression Grading: We categorized all MDD patients into two groups based on their symptom severity as assessed by the 17-item Hamilton Depression Rating Scale (HAMD$_{17}$) scores. According to an established clinical criteria\cite{boessen2013comparing}, 138 patients with baseline HAMD$_{17}<$ 17 before treatment were categorized as a mild depression group while 127 as moderate to severe depression group. We tested our model for classifying these two groups.
For Sex Classification: We also tested our model on discriminating sex in MDD patients (103 females and 162 males).

\subsubsection{HBN}
The HBN \cite{alexander2017open}, created by the Child Mind Institute, is an open resource for transdiagnostic research on children's mental health and learning disorders. It includes rich multimodal data such as structural and functional MRI, EEG, cognitive-behavioral assessments, and genomic data. The dataset covers a range of pediatric mental health conditions like ADHD, autism, depression, and anxiety. HBN's transdiagnostic approach enables researchers to explore commonalities and differences across diagnoses. 
In our study, we used fMRI and EEG time series data from the HBN dataset. For the fMRI data, HBN includes 2,282 subjects, with each subject having two fMRI runs. For the EEG data, HBN contains eye-open EEG time series from 1,594 subjects and eye-closed EEG time series from 1,744 subjects. To match subjects with both fMRI and EEG data, we performed a selection process and identified 1,029 subjects who had both fMRI and EEG data available.
In subjects who have undergone modality matching, we conducted two downstream tasks: the identification of MDD and the identification of ASD. And we quantified the labeled data for both disorders. Specifically, among the subjects with MDD, 119 were diagnosed as healthy, while 77 were diagnosed as MDD patients. In the case of ASD subjects, there were 119 healthy individuals and 138 diagnosed with ASD.

\begin{table}[h]
\footnotesize
\renewcommand\arraystretch{0.6}
\setlength{\tabcolsep}{4pt}
\centering
\caption{Statistical summary of subjects and runs across different modalities for five datasets for pre-training. For the embarc and hbn datasets, each subject's fMRI scans consist of two runs, while each EEG dataset includes scans for both eye-open and eye-closed conditions (eye-open / eye-closed).}
\label{exp_tb_data}
\begin{tabular}{cccccccc}
\toprule
\multirow{2}{*}{Datasets} & \multirow{2}{*}{ADHD-200} & \multirow{2}{*}{ABIDE I} & \multirow{2}{*}{ABIDE II} & \multicolumn{2}{c}{EMBARC} & \multicolumn{2}{c}{HBN} \\ \cmidrule(lr){5-6} \cmidrule(lr){7-8}
 &fMRI &fMRI &fMRI & fMRI  & EEG  & fMRI  & EEG  \\ \midrule
Subjects  & 830 & 1,102 & 990 & 324 & 308/308 & 2,282 & 1,594/1,744 \\ 
Runs     & 1086 & 1,153 & 1,314 & 648 & 616 & 4,564 & 3,338 \\ \midrule
Paired Subjects    & - & - & - & \multicolumn{2}{c}{308} & \multicolumn{2}{c}{1,029} \\ 
Paired Runs    & - & - & - & \multicolumn{2}{c}{616} & \multicolumn{2}{c}{2,058} \\  
Cross Pairs    & - & - & - & \multicolumn{2}{c}{1,232} & \multicolumn{2}{c}{4,116} \\ \bottomrule
\end{tabular}
\end{table}

\begin{figure}[htbp]
	\centering
	\includegraphics[width=\linewidth]{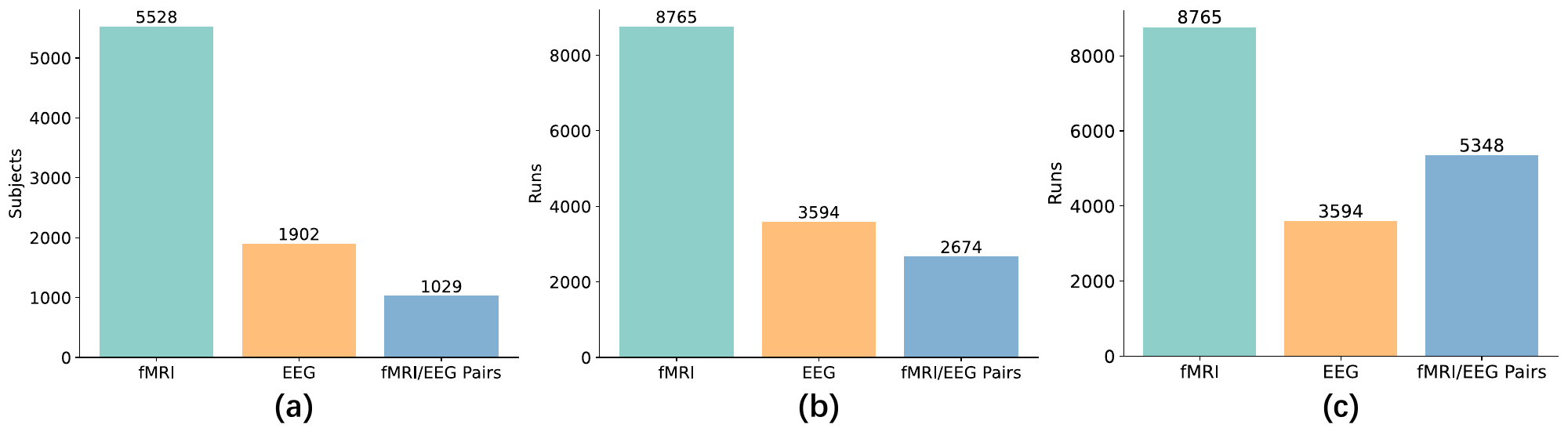}
	\caption{Statistics on the number of runs and pairs for different subjects across all datasets for both fMRI and EEG modalities. Fig.(a) shows the total statistics of the number of subjects for fMRI, EEG, and their pairs across all datasets. Fig.(b) displays the comprehensive statistics of the number of runs for these three modalities. Fig.(c) presents the overall runs for the three modalities, where the number of pairs is calculated through cross-matching between different runs of fMRI and EEG.}
	\label{exp_fg_runs}
\end{figure}

\subsection{Unified Pre-training Dataset Construction}
As shown in Table \ref{exp_tb_data}, we compiled statistics on the number of fMRI and EEG scans across the five datasets. The ADHD-200 and ABIDE datasets contain only fMRI data, whereas the EMBARC and HBN datasets provide both fMRI and EEG data. In these multimodal datasets, fMRI data for each subject includes two runs, and EEG data comprises two runs corresponding to eye-open and eye-closed conditions.
However, due to missing modalities in several subjects, the number of usable fMRI-EEG pairs for pre-training is relatively limited, potentially compromising the effectiveness of pre-training. To mitigate this issue, we applied a cross-matching strategy, pairing different runs of fMRI and EEG from the same subject. This approach allows each subject to generate up to $2^2$ pairs, thereby significantly increasing the number of available pre-training samples.

As shown in Figure \ref{exp_fg_runs}, we developed three pre-training datasets for fMRI, EEG, and fMRI/EEG pairs by integrating data from all five original datasets. The fMRI dataset has a relatively large volume, whereas the EEG dataset is smaller in size. However, by employing our cross-matching strategy, we were able to achieve a substantial number of fMRI/EEG paired samples as well.


\subsection{Data preprocessing}
All rs-fMRI data utilized in this study were preprocessed using the fMRIPrep pipeline\cite{esteban2019fmriprep}. The T1-weighted image underwent intensity non-uniformity correction and skull stripping, followed by spatial normalization through nonlinear registration\cite{avants2008symmetric}. Brain features were segmented using FSL, and fieldmap information was used for distortion correction. The BOLD reference was transformed to the T1-weighted image, addressing remaining distortion with nine degrees of freedom. Head-motion parameters were estimated, and BOLD signals were corrected and resampled into standard space. ICA-AROMA was applied for artifact removal\cite{pruim2015ica}. 
Regional time series were then extracted from the preprocessed rs-fMRI using the Schaefer atlas of 100 ROIs, followed by calculating functional connectivity using Pearson's correlation. Based on these connectivity measures, we constructed brain network graphs for the subsequent analysis.

In the EMBARC and HBN datasets, rs-EEG signals were preprocessed using an automated artefact rejection pipeline \cite{delorme2004eeglab}, involving removing line noise, drifts, and artefacts, rejecting bad epochs and channels, interpolating bad channels, and re-referencing to the common average. The preprocessed EEG were filtered into four bands: theta (4-7 Hz), alpha (8-12 Hz), beta (13-30 Hz), and gamma (31-50 Hz). EEG source localization and power envelope connectivity were calculated using the approach described in our previous study \cite{zhang2021identification}. The same brain atlas as used for fMRI analysis was adopted for calculating ROI-level EEG time series and connectivity.
For EEG signals in each band, we computed connectivity and tested each band independently. In line with previous studies indicating the alpha band's relevance in ADHD \cite{koehler2009increased}, ASD \cite{dickinson2018peak}, and MDD \cite{tement2016eeg}, our analysis focused on the alpha band due to its documented significance in these disorders. Consequently, all EEG analyses in this study centered on the alpha band signals.


\section{Experiments}
\label{sec:experiment}

\subsection{Implementation details}
Our experimental analyses were implemented using the PyTorch framework and trained on an Nvidia RTX 4090 GPU. The total number of training epochs was set to 50. We employed dynamic learning rates and Adam optimizer for a better model training. We conducted 10 runs of 10-fold cross validation. For each run, we adopted different random seeds to construct training, validation and testing sets. We fixed random seeds for all the models but adopted random seeds to split dataset in each run. 
For pre-training, ensuring the effectiveness of contrastive learning requires both a substantial data volume and a sufficiently large batch size. Consequently, we set the batch size to 128 during this phase. For fine-tuning, however, due to the limited availability of labeled data for specific tasks, we reduced the training batch size to 32. The learning rate was set to 0.0005.

\subsection{Overall comparison with other methods}
We compared our model with a wide array of state-of-the-art approaches on the ADHD-200, ABIDE I, ABIDE II, EMBARC and HBN datasets for various classification tasks. This encompasses both traditional machine learning methods (SVM, Random Forest, MLP) and advanced deep learning models (BrainNetCNN\cite{kawahara2017brainnetcnn}, BrainGNN\cite{li2021braingnn}, GCN\cite{kipf2016semi}, GAT\cite{velivckovic2017graph}, Graph Transformer\cite{yun2019graph}, FBNetGen\cite{kan2022fbnetgen}, BNT\cite{kan2022brain}, Dynamic BNT\cite{kan2023dynamic}). 

As shown in Table \ref{tb_abide}, our method outperformed all other compared approaches for ADHD and ASD diagnoses, as evaluated by commonly used classification metrics: AUROC, Accuracy (Acc), Recall, and Precision. For the single-modality ADHD-200, ABIDE I and ABIDE II datasets, we leveraged only the proposed cross-domain self-supervised loss for pretraining and exploring interaction relationships between different domains, resulting in a significant improvement compared to models without pretraining. As shown in Table \ref{tb_embarc} and \ref{tb_hbn}, we also evaluated our method on the EMBARC and HBN datasets, which incorporated both fMRI and EEG data. We conducted both cross-domain and cross-modality self-supervised loss for the pre-training. Our model also outperformed the comparative methods and achieved significant advancements beyond the baseline method without pretraining, across both depression grading and sex classification tasks. In summary, our baseline method without self-supervised pre-training achieved comparable performance to the best existing methods. More importantly, the incorporation of cross-domain and cross-modality constraints through our approach surpasses the classification performance of previous methods.

\begin{table*}[htbp]
	\tiny
	\renewcommand\arraystretch{1.0}
	\setlength{\tabcolsep}{1pt}
	\centering
	\caption{Quantitative comparison of \textbf{ADHD and ASD prediction} on the \textbf{ADHD-200}, \textbf{ABIDE I} and \textbf{ABIDE II} datasets between our model and other existing approaches. 'PT' means pre-training.}
	\begin{tabular}{c|cccc|cccc|cccc}
		\toprule
		& \multicolumn{4}{c|}{ADHD-200} & \multicolumn{4}{c|}{ABIDE I} & \multicolumn{4}{c}{ABIDE II} \\
		\midrule
		Method  & AUROC    & Acc    & Recall   & Precision  & AUROC    & Acc    & Recall   & Precision  & AUROC    & Acc    & Recall   & Precision   \\
		\midrule
		SVM    &59.6$\pm$3.7    &60.3$\pm$3.4    &58.7$\pm$4.3     &61.6$\pm$3.6 &60.8$\pm$3.6    &62.4$\pm$2.3    &58.9$\pm$3.9    &61.5$\pm$4.2   &61.2$\pm$3.2   &60.8$\pm$4.6    &60.4$\pm$3.8     &62.5$\pm$2.4  \\
		Random Forest    &62.7$\pm$2.1    &63.8$\pm$2.9    &62.8$\pm$3.1     &60.4$\pm$3.7     &61.4$\pm$2.6    &62.8$\pm$3.5   &63.1$\pm$2.4    &61.6$\pm$4.2  &62.6$\pm$2.1  &63.1$\pm$2.5    &61.5$\pm$3.6     &62.9$\pm$2.1   \\
		MLP   &61.4$\pm$3.5    &62.9$\pm$3.8    &62.2$\pm$3.7     &61.8$\pm$4.2     &62.3$\pm$3.4    &61.7$\pm$3.1   &62.6$\pm$2.8    &60.9$\pm$3.5  &63.3$\pm$2.4   &62.5$\pm$3.7    &63.7$\pm$3.3     &61.7$\pm$4.3   \\
		\midrule
		BrainNetCNN  \cite{kawahara2017brainnetcnn}   &61.1$\pm$4.5    &60.6$\pm$5.3    &59.4$\pm$4.9     &61.7$\pm$4.2    &61.4$\pm$2.7   &59.8$\pm$4.1    &60.2$\pm$3.4    &62.1$\pm$2.5  &62.9$\pm$3.3   &63.5$\pm$2.5    &60.3$\pm$3.1     &61.4$\pm$2.6  \\
		BrainGNN  \cite{li2021braingnn}  &64.7$\pm$2.8   &63.2$\pm$1.4    &65.4$\pm$2.4     &63.5$\pm$3.6   &65.3$\pm$2.3   &63.9$\pm$3.5    &65.3$\pm$2.2     &63.2$\pm$3.9  &66.2$\pm$3.2   &65.1$\pm$2.7    &63.8$\pm$4.3     &65.4$\pm$3.2    \\
		GCN  \cite{kipf2016semi}  &65.8$\pm$3.5    &66.1$\pm$1.8    &64.4$\pm$4.5     &67.1$\pm$5.1   &64.6$\pm$2.3    &65.8$\pm$2.7    &64.3$\pm$3.3     &63.8$\pm$4.3   &65.7$\pm$2.6    &65.8$\pm$3.7    &64.6$\pm$3.6     &65.2$\pm$2.3   \\
		GAT \cite{velivckovic2017graph}  &66.2$\pm$2.9    &67.4$\pm$2.3    &67.6$\pm$3.3     &65.8$\pm$2.5     &66.2$\pm$1.6    &65.6$\pm$2.8    &66.3$\pm$2.3     &64.4$\pm$3.5    &64.2$\pm$4.1    &66.5$\pm$2.5    &66.2$\pm$3.2     &65.7$\pm$2.4  \\
		G Transformer  \cite{yun2019graph}   &66.5$\pm$2.7    &65.9$\pm$3.1    &67.7$\pm$2.5     &66.3$\pm$3.6  &65.3$\pm$3.6    &63.8$\pm$4.2   &65.6$\pm$2.7     &64.2$\pm$3.3  &66.5$\pm$2.5    &65.8$\pm$3.8    &65.3$\pm$3.4     &67.1$\pm$2.9   \\
		FBNetGen \cite{kan2022fbnetgen}  &67.1$\pm$2.4    &66.4$\pm$3.1  &66.8$\pm$3.9     &68.2$\pm$2.7   &68.2$\pm$2.7    &66.7$\pm$3.1  &67.1$\pm$3.6     &65.4$\pm$4.3  &67.5$\pm$1.6    &67.8$\pm$2.8  &66.3$\pm$3.1     &64.2$\pm$4.2  \\
		BNT \cite{kan2022brain}   &67.7$\pm$2.9    &68.7$\pm$2.6    &66.9$\pm$4.3     &65.6$\pm$3.1    &68.3$\pm$2.8    &68.1$\pm$3.2    &66.7$\pm$3.7     &65.3$\pm$3.4   &69.1$\pm$2.5    &68.9$\pm$3.8    &66.6$\pm$3.4     &68.8$\pm$2.9  \\
		Dynamic BNT \cite{kan2023dynamic}   &68.1$\pm$3.2    &68.8$\pm$2.5    &67.5$\pm$3.5     &66.8$\pm$2.9  &67.7$\pm$3.1    &68.2$\pm$2.9    &65.5$\pm$3.7     &65.8$\pm$4.3  &68.5$\pm$2.1    &70.5$\pm$2.4    &67.4$\pm$2.8     &68.3$\pm$3.3   \\
		\midrule
		\textbf{MCSP w/ PT} &79.6$\pm$2.1    &71.7$\pm$3.1   &70.4$\pm$3.3     &67.9$\pm$2.6   &70.2$\pm$2.7    &70.8$\pm$2.4    &69.5$\pm$3.1     &67.2$\pm$2.8   &71.5$\pm$2.3    &72.6$\pm$1.8    &68.2$\pm$2.4     &69.8$\pm$3.5   \\
		\bottomrule
	\end{tabular}
	\label{tb_abide}
\end{table*}

\begin{table*}[htbp]
	\scriptsize
	\renewcommand\arraystretch{1.}
	\setlength{\tabcolsep}{3pt}
	\centering
	\caption{Quantitative comparison of the \textbf{Depression Grading} and \textbf{MDD Sex Classification} tasks on \textbf{EMBARC} dataset between our model and other existing approaches.}
	\begin{tabular}{c|cccc|cccc}
		\toprule
		& \multicolumn{4}{c|}{Depression Grading}  & \multicolumn{4}{c}{Sex Classification} \\
		\midrule
		Method  & AUROC    & Acc    & Recall   & Precision    & AUROC    & Acc    & Recall   & Precision    \\
		\midrule
		SVM           &60.4$\pm$4.2   &62.3$\pm$3.9    &61.5$\pm$3.1    &60.7$\pm$3.6   &63.6$\pm$3.1    &66.7$\pm$2.8   &63.8$\pm$4.3    &64.3$\pm$3.2  \\
		Random Forest       &62.1$\pm$1.9   &63.6$\pm$2.7    &64.3$\pm$3.5    &62.6$\pm$2.8    &64.8$\pm$2.7    &67.4$\pm$3.2   &65.9$\pm$2.5     &67.3$\pm$2.3   \\
		MLP      &61.8$\pm$3.5    &64.6$\pm$2.7    &61.9$\pm$3.2     &62.5$\pm$3.7    &66.3$\pm$3.5    &67.6$\pm$2.8   &64.8$\pm$3.1     &65.2$\pm$3.5   \\
		\midrule
		BrainNetCNN  \cite{kawahara2017brainnetcnn}    &63.3$\pm$2.7    &64.2$\pm$2.2    &63.5$\pm$4.3     &62.7$\pm$3.1    &65.9$\pm$3.8    &67.4$\pm$3.3   &66.7$\pm$2.9     &62.9$\pm$2.4 \\
		BrainGNN  \cite{li2021braingnn}    &61.4$\pm$3.0   &64.2$\pm$1.8    &62.3$\pm$2.7     &65.3$\pm$2.3     &63.3$\pm$2.9   &65.1$\pm$2.2   &62.8$\pm$3.5     &64.3$\pm$2.6  \\
		GCN  \cite{kipf2016semi}    &64.3$\pm$3.4    &65.7$\pm$2.7    &66.5$\pm$4.1     &65.2$\pm$3.8     &65.3$\pm$2.6    &68.4$\pm$2.1   &66.2$\pm$2.5     &67.3$\pm$3.8    \\
		GAT  \cite{velivckovic2017graph}    &65.3$\pm$2.9    &67.0$\pm$3.2    &65.8$\pm$3.4     &65.4$\pm$2.7    &67.3$\pm$3.5    &68.6$\pm$2.9   &67.3$\pm$3.2    &66.8$\pm$3.4  \\
		G Transformer  \cite{yun2019graph}   &65.7$\pm$3.2    &66.4$\pm$3.7    &64.2$\pm$2.9     &65.7$\pm$3.2    &66.3$\pm$2.7    &70.6$\pm$2.2   &66.3$\pm$4.2     &68.4$\pm$3.7  \\
		FBNetGen \cite{kan2022fbnetgen}    &64.4$\pm$3.7    &67.7$\pm$2.5  &65.2$\pm$3.6     &66.7$\pm$4.2     &67.1$\pm$2.1    &71.2$\pm$2.8  &66.7$\pm$4.5    &64.9$\pm$3.2  \\
		BNT \cite{kan2022brain}    &65.1$\pm$2.7    &67.6$\pm$3.2    &66.8$\pm$3.8     &67.3$\pm$4.2   &70.3$\pm$2.9    &73.8$\pm$2.7   &68.6$\pm$3.5     &70.9$\pm$4.2   \\
		Dynamic BNT  \cite{kan2023dynamic}    &66.6$\pm$3.1    &68.2$\pm$3.4    &66.4$\pm$4.1     &64.6$\pm$3.5      &69.2$\pm$2.7    &72.4$\pm$3.7    &67.8$\pm$2.3     &69.3$\pm$3.4   \\
		\midrule
		\textbf{MCSP w/o PT} &66.3$\pm$2.9    &67.8$\pm$2.4    &67.3$\pm$2.2     &65.6$\pm$2.8    &68.5$\pm$2.4    &72.3$\pm$3.1    &67.9$\pm$3.7     &70.6$\pm$3.2  \\
		\textbf{MCSP w/ PT} &69.5$\pm$2.5    &71.9$\pm$2.7    &71.1$\pm$3.3     &69.4$\pm$2.8     &72.6$\pm$2.9    &77.2$\pm$3.1    &73.6$\pm$3.4     &75.7$\pm$2.2 \\
		\bottomrule
	\end{tabular}
	\label{tb_embarc}
\end{table*}

\begin{table*}[htbp]
	\scriptsize
	\renewcommand\arraystretch{1.}
	\setlength{\tabcolsep}{3pt}
	\centering
	\caption{Quantitative comparison of the \textbf{MDD Diagnosis} and \textbf{ASD Diagnosis} tasks on \textbf{HBN} dataset between our model and other existing approaches.}
	\begin{tabular}{c|cccc|cccc}
		\toprule
		& \multicolumn{4}{c|}{MDD Diagnosis}  & \multicolumn{4}{c}{ASD Diagnosis} \\
		\midrule
		Method  & AUROC    & Acc    & Recall   & Precision    & AUROC    & Acc    & Recall   & Precision    \\
		\midrule
		SVM     &67.2$\pm$3.4 &69.7$\pm$2.7 &60.5$\pm$2.7 &66.9$\pm$3.2 &59.6$\pm$2.8 &64.3$\pm$2.5 &70.6$\pm$3.3 &66.6$\pm$2.9 \\
		Random Forest   &69.3$\pm$1.8 &70.3$\pm$2.5 &63.6$\pm$3.6 &68.1$\pm$2.2 &61.4$\pm$1.9 &67.9$\pm$2.6 &71.5$\pm$2.5 & 67.6$\pm$3.6  \\
		MLP     &67.7$\pm$3.2 &70.8$\pm$2.9 &64.1$\pm$3.6 &68.5$\pm$2.7 &60.7$\pm$3.4 &66.2$\pm$2.3 &69.6$\pm$4.2 &71.9$\pm$2.2  \\
		\midrule
		BrainNetCNN  \cite{kawahara2017brainnetcnn}    &69.4$\pm$2.7 &71.7$\pm$3.5 &62.8$\pm$2.1 &68.4$\pm$2.8 &60.4$\pm$1.5 &66.8$\pm$2.6 &71.4$\pm$2.6 &67.8$\pm$3.7  \\
		BrainGNN  \cite{li2021braingnn}    &68.4$\pm$2.3 &73.2$\pm$2.1 &63.7$\pm$2.4 &69.3$\pm$3.1 &62.4$\pm$2.1 & 67.9$\pm$2.6  &70.8$\pm$3.5 &67.3$\pm$2.7  \\
		GCN  \cite{kipf2016semi}    &70.9$\pm$1.8 &73.6$\pm$2.4 &68.1$\pm$2.3 &66.7$\pm$3.3 &63.3$\pm$2.5 &68.7$\pm$3.2 &73.7$\pm$1.8 &69.3$\pm$2.1  \\
		GAT  \cite{velivckovic2017graph}    &71.4$\pm$2.5 &75.4$\pm$2.2 &70.7$\pm$3.1 &67.2$\pm$2.9 &63.7$\pm$1.7 &69.3$\pm$2.3 &77.4$\pm$3.6 &67.7$\pm$4.2  \\
		G Transformer  \cite{yun2019graph}   &72.2$\pm$2.5 &74.9$\pm$3.4 &64.6$\pm$3.4 &71.7$\pm$2.2 &63.8$\pm$2.1 &71.3$\pm$1.5 &75.6$\pm$2.2 &69.1$\pm$3.6  \\
		FBNetGen \cite{kan2022fbnetgen}    &71.3$\pm$2.6 &75.3$\pm$3.4 &65.3$\pm$2.3 &70.5$\pm$2.9 &64.5$\pm$2.5 &68.1$\pm$2.1 &76.7$\pm$2.6 &68.8$\pm$3.3 \\
		BNT \cite{kan2022brain}   &73.7$\pm$2.7 &77.4$\pm$1.5 &67.4$\pm$3.1 &72.8$\pm$2.3 &65.5$\pm$2.2 &69.7$\pm$4.2 &79.5$\pm$1.9 &70.4$\pm$3.4   \\
		Dynamic BNT  \cite{kan2023dynamic}    &73.4$\pm$2.8 &76.3$\pm$1.4 &72.3$\pm$3.3 &68.7$\pm$2.5 &66.6$\pm$3.4 &70.7$\pm$2.2 &77.8$\pm$1.8 & 73.3$\pm$3.4 \\
		\midrule
		\textbf{MCSP w/o PT}    &74.3$\pm$1.7 &76.9$\pm$3.5 & 68.6$\pm$2.4 & 72.6$\pm$3.5  &68.2$\pm$1.9 &71.4$\pm$2.6  &73.7$\pm$4.3  &77.1$\pm$2.6   \\
		\textbf{MCSP w/ PT}    &77.6$\pm$2.8 &80.3$\pm$1.7  &69.5$\pm$3.4 &75.3$\pm$2.2 &71.6$\pm$1.6 &74.8$\pm$3.6     &80.3$\pm$2.1  &74.8$\pm$3.6   \\
		\bottomrule
	\end{tabular}
	\label{tb_hbn}
\end{table*}


\subsection{Ablation analysis of cross domains and modalities}
To investigate the influence of different domains and modalities on the model, we conducted comprehensive ablation experiments both quantitatively and qualitatively. As shown in Table \ref{tb_abla_all}, for fMRI, the spatial domain outperformed the temporal and frequency domains by approximately 4\% in terms of both AUROC and Accuracy metrics. A modest enhancement in performance was further observed by integrating all the three domains. As for EEG, the performance across the three domains is relatively similar, displaying a marginal decrease in performance when compared to those of fMRI. While integrating fMRI and EEG yielded improvements, the enhancements is not significant.  A likely explanation for the minimal improvement post-fusion could be the simplistic approach taken towards integrating features from the three domains of fMRI or EEG, without a focused strategy to uncover and exploit latent interactions among these domains. Therefore, our proposed cross-domain self-supervised loss serves to address this gap, explicitly modeling inter-domain relationships without adding additional parameter burdens to the model.

\begin{table}[]
	\footnotesize
	\renewcommand\arraystretch{0.5}
	\setlength{\tabcolsep}{5pt}
	\centering
	\caption{Overall ablation study of each domain for different modalities on \textbf{EMBARC} for Depression Grading task. Note that all the methods employed simple fusion of domains and modalities without constraints and pre-training between different domains and modalities.}
	\begin{tabular}{c|c|cccc}
		\toprule

		Modality        &Domain      & AUROC    & Acc    & Recall    & Precision      \\
		\midrule
		fMRI        &Spatial         &64.4$\pm$2.7    &65.9$\pm$2.3   &64.6$\pm$2.8     &65.3$\pm$3.1   \\

		fMRI           &Temporal     &60.3$\pm$3.2    &62.3$\pm$3.2    &57.4$\pm$3.6     &62.8$\pm$2.5  \\
		fMRI           &Frequency    &59.8$\pm$3.6    &61.1$\pm$3.2    &60.8$\pm$2.8     &58.6$\pm$2.4  \\
		\midrule
		fMRI           &All    &65.4$\pm$2.3    &65.7$\pm$3.2    &64.2$\pm$2.9     &65.8$\pm$2.8   \\
		\midrule
		\midrule
		Modality        &Domain                     & AUROC    & Acc    & Recall    & Precision    \\
		\midrule
		EEG      &Spatial          &65.2$\pm$3.2    &66.4$\pm$2.8   &65.8$\pm$3.4     &64.7$\pm$3.6   \\

		EEG        &Temporal         &64.6$\pm$3.5    &65.4$\pm$2.9   &62.6$\pm$2.8    &64.3$\pm$3.3   \\
		EEG           &Frequency       &62.8$\pm$2.7    &65.5$\pm$1.9   &65.6$\pm$3.2    &62.8$\pm$2.3   \\
		\midrule
		EEG         &All     &64.2$\pm$2.7   &67.5$\pm$2.8   &65.3$\pm$4.2     &65.9$\pm$3.1   \\
		\midrule
		\midrule
		Modality        &Domain                    & AUROC    & Acc    & Recall    & Precision   \\
		\midrule
		fMRI + EEG    &Spatial        &66.3$\pm$3.5    &67.2$\pm$2.2   &64.3$\pm$3.8     &66.7$\pm$2.5  \\
		
		fMRI + EEG    &Temporal   &64.2$\pm$4.4   &67.1$\pm$2.4   &63.7$\pm$3.2     &66.3$\pm$1.6   \\
		fMRI + EEG    &Frequency    &63.6$\pm$3.4    &65.3$\pm$3.6   &63.8$\pm$2.8     &65.4$\pm$4.6   \\
		\midrule
		fMRI + EEG    &All   &66.3$\pm$2.9    &67.8$\pm$2.4    &67.3$\pm$2.2     &65.6$\pm$2.8  \\

		\bottomrule
	\end{tabular}
	\label{tb_abla_all}
\end{table}

\subsection{Brain biomarker interpretation}

\begin{figure}[htbp]
	\centering
	\includegraphics[width=\linewidth]{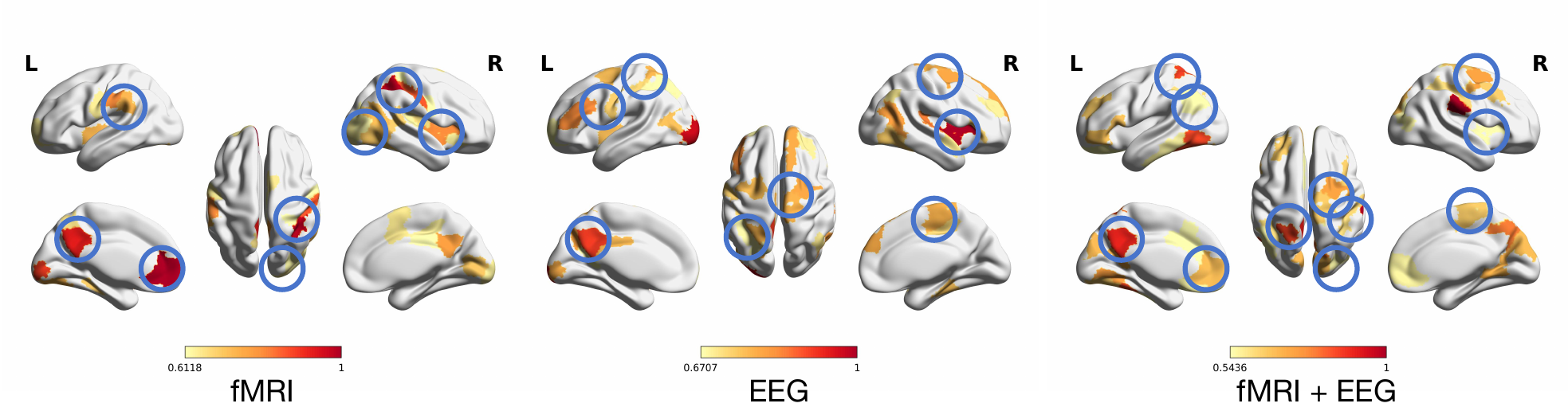}
	\caption{Visual ablation study for different modalities on the EMBARC dataset for Depression Grading. For each modality, we combine all the three domains together for the training. We visualized the top 10 most important brain regions and provided the connections between these significant regions. The blue circles represent the important brain regions jointly selected by different modalities.}
	\label{depress}
\end{figure}


\begin{figure}[htbp]
	\centering
	\includegraphics[width=\linewidth]{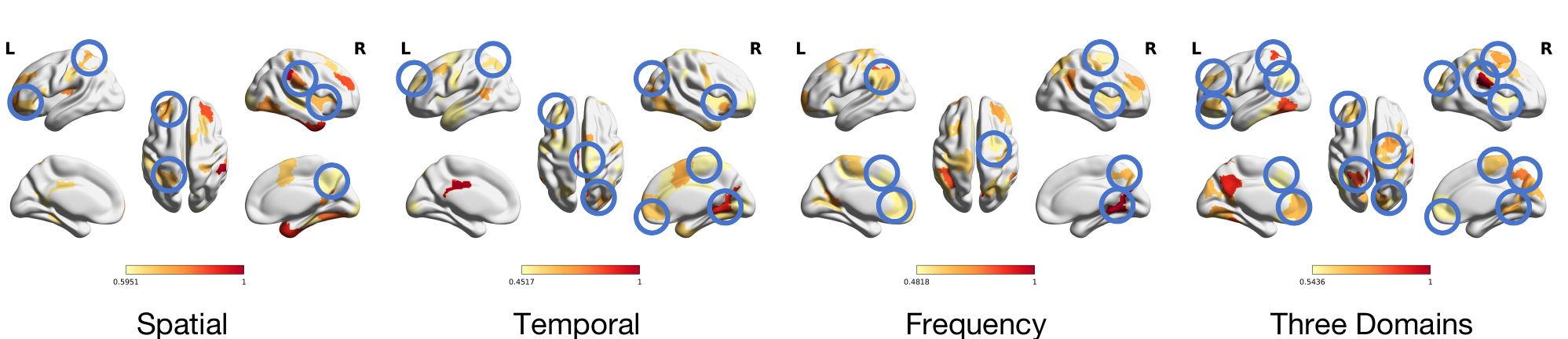}
	\caption{Visual ablation study for different domains on the EMBARC dataset for Depression Grading task. For each domain, we integrate both fMRI and EEG modalities together for the training. The blue circles represent the important brain ROIs jointly selected by different domains.}
	\label{domain}
\end{figure}

As shown in Figs. \ref{depress} and \ref{domain}, We ranked the brain regions according to their importance scores and visualized the top 10 most important brain ROIs identified from each domain and modality for the Depression Grading task on the EMBARC dataset, providing an interpretation of brain areas associated with depression severity. Our results indicate that the Lingual Cortex, Superior Parietal Cortex, Superior Frontal Cortex and Inferior Temporal Cortex are most important for predicting depression severity. These findings echo the results of numerous previous studies. For instance, the Lingual Cortex's dysfunction has been noted for its strong linkage to MDD severity and related anxiety, emphasizing its critical role in mood regulation \cite{couvy2018lingual}. The Superior Parietal and Frontal Cortex, involved in spatial attention and sensory integration, has been implicated in emotional processing, particularly in regulating attention to emotional stimuli. In addition, the Inferior Temporal Cortex, associated with visual object recognition and emotional valence, is intricately connected to the limbic system, influencing emotion regulation and cognitive functions. Disruptions in neural circuits involving these regions have suggested to underlie the emotionl and cognitive symptoms observed in MDD. Studies by Greicius et al. \cite{greicius2007resting} and Fox et al. \cite{fox2014resting} have highlighted the relevance of these regions in resting-state functional connectivity and their involvement in emotional regulation. Furthermore, alterations in the Superior Parietal Cortex and Inferior Temporal Cortex have been associated with cognitive deficits and emotional dysregulation in MDD, as discussed by Price et al. \cite{price2010neurocircuitry} and Mayberg et al. \cite{mayberg2003modulating}. These insights into the functional and structural importance of these cortical regions in MDD underscore their potential as targets for therapeutic strategies aimed at modulating the underlying neurobiological mechanisms of the disorder.


\subsection{Ablation study on the proposed self-supervised loss}
As shown in Table \ref{tb_loss}, we investigated the effects of the proposed cross-domain and cross-modal self-supervised loss functions, comparing a baseline model that used only cross-entropy loss for classification. The incorporation of self-supervised pretraining, leveraging both CD-SSL and CM-SSL markedly enhanced the model's performance. It is worth noting that these proposed loss functions not only effectively serve as the basis for our pretraining procedure but also demonstrate comparable efficiency in improving model outcomes. A key advantage of employing this pretraining approach is its versatility, allowing the pretrained model to be adeptly adapted and applied to a variety of downstream tasks as needed.



\begin{table*}[]
	\scriptsize
	\renewcommand\arraystretch{0.8}
	\setlength{\tabcolsep}{2pt}
	\centering
	\caption{Overall ablation study of the proposed loss functions on each domain for different modalities on \textbf{EMBARC} for \textbf{Depression Grading} and \textbf{Sex Classification} tasks.}
	\begin{tabular}{c|c|cccc|cccc}
		\toprule
		& & \multicolumn{4}{c|}{Depression Grading} & \multicolumn{4}{c}{Sex Classification} \\
		\midrule
		Pre-trained   &Loss      & AUROC    & Acc    & Recall    & Precision     & AUROC    & Acc    & Recall    & Precision     \\
		\midrule
		\XSolidBrush    &+ CE       &66.3$\pm$2.9    &67.8$\pm$2.4    &67.3$\pm$2.2     &65.6$\pm$2.8   &68.5$\pm$2.4    &72.3$\pm$3.1    &67.9$\pm$3.7     &70.6$\pm$3.2   \\
		\Checkmark    &+ CD-SSL   &67.5$\pm$2.8    &68.8$\pm$3.2    &67.7$\pm$3.4     &68.1$\pm$2.6   &71.2$\pm$2.6    &73.9$\pm$3.2    &71.6$\pm$3.7     &72.4$\pm$2.5  \\
		\Checkmark    &+ CM-SSL   &69.5$\pm$2.5    &71.9$\pm$2.7    &71.1$\pm$3.3     &69.4$\pm$2.8     &72.6$\pm$2.9    &77.2$\pm$3.1    &73.6$\pm$3.4     &75.7$\pm$2.2   \\

		\bottomrule
	\end{tabular}
	\label{tb_loss}
\end{table*}

\subsection{Analysis of the potential universal pre-training abilities}
\begin{figure}[htbp]
	\centering
	\includegraphics[width=\linewidth]{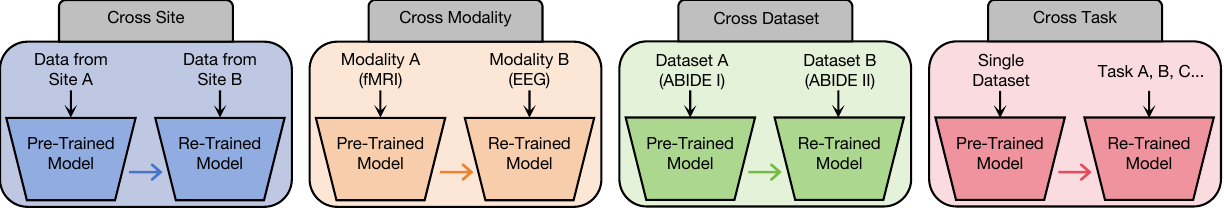}
	\caption{Different scenarios of the universal pre-training.}
	\label{function}
\end{figure}

As illustrated in Fig. \ref{function}, our model demonstrates the capability of universal pretraining on fMRI and EEG across four scenarios. \textbf{1) Cross Modality}: The model was pretrained on fMRI and fine-tuned on EEG, and vice versa. The universal pretraining strategy, utilizing the proposed cross-domain loss, yielded significant improvements in both AUROC and Accuracy. \textbf{2) Cross Dataset}: By pre-training on one dataset (e.g., ABIDE I) and fine-tuning on another (e.g., ABIDE II), the model demonstrates its adaptability, with the cross-domain loss contributing to superior performance on the new dataset. \textbf{3) Cross Task}: Employed in a task-agnostic manner, the model was first pretrained and then fine-tuned for specific tasks, such as MDD or ASD  diagnosis within the HBN dataset. This approach resulted in notable performance improvements, evidencing the model's cross-task adaptability. \textbf{4) Cross Site}: Pre-training on data from one site (e.g., NYU) and fine-tuning on data from another (e.g., PKU), as verified using the ADHD-200 dataset, our model effectively bridged site-specific data variations, extracting valuable information even in the absence of site-specific labels. The results summarized in Table \ref{tb_function} validate the generality of our model and highlight the advantages of universal pre-training in enhancing versatility and adaptability to varied data sources, tasks, and collection sites. Employing the self-supervised loss in both pre-training and fine-tuning phases, our method, as evidenced by Table \ref{tb_function}, achieved substantial improvements through transfer learning. The remarkable gains observed on the HBN dataset can be ascribed to its multi-modality composition, where our model capitalizes on both cross-domain and cross-modality constraints, effectively synthesizing information across different modalities. This resulted in more pronounced improvements when compared to single-modality datasets like ADHD-200 and ABIDE.

\begin{table}[ht]
	\footnotesize
	\renewcommand\arraystretch{0.8}
	\setlength{\tabcolsep}{8pt}
	\centering
	\caption{Experiments for demonstrating the universal pre-training abilities on different scenarios. Note that 'Both' means pre-training on both fMRI and EEG modalities. All improvements in performance are compared against the model without pre-training and the proposed loss functions.}
	\begin{tabular}{c|c|c|c|cc}
		\toprule
		Scenario        &Dataset    &Pre-train        &Fine-tune      & AUROC    & Acc         \\
		\midrule
		Cross Modality  &HBN   &fMRI    &EEG    &2\%$\uparrow$    &2\%$\uparrow$      \\
		Cross Modality    &HBN    &EEG    &fMRI   &1\%$\uparrow$    &2\%$\uparrow$  \\
		\midrule 
  Cross Modality  &HBN   &Both    &fMRI    &2\%$\uparrow$    &2\%$\uparrow$      \\
		Cross Modality    &HBN    &Both    &EEG   &3\%$\uparrow$    &2\%$\uparrow$  \\
		\midrule  
		Cross Dataset   &ABIDE  &ABIDE I    &ABIDE II   &3\%$\uparrow$    &3\%$\uparrow$      \\
		Cross Dataset    &ABIDE   &ABIDE II    &ABIDE I   &2\%$\uparrow$    &3\%$\uparrow$      \\
		\midrule 
		Cross Task    &HBN   &HBN    &MDD   &3\%$\uparrow$    &4\%$\uparrow$      \\
		Cross Task     &HBN  &HBN    &ASD   &2\%$\uparrow$    &3\%$\uparrow$    \\
		\midrule 
		Cross Site     &ADHD200   &NYU    &PKU  &2\%$\uparrow$    &2\%$\uparrow$     \\
		Cross Site     &ADHD200    &PKU    &NYU  &1\%$\uparrow$    &2\%$\uparrow$     \\
		\bottomrule
	\end{tabular}
	\label{tb_function}
\end{table}

		

\subsection{Cross-model knowledge distillation across domains}

A notable application of our model lies in its adeptness at facilitating cross-model distillation across diverse domains. Initially, we constructed a comprehensive teacher model by integrating data from spatial, temporal, and spectral domains, serving as a repository of consolidated knowledge. We then employed distillation techniques to transfer the insights gathered by this teacher model to smaller, domain-specific student models. This process significantly enhances their performance in specialized tasks. As delineated in Table \ref{tb_dis}, the teacher model's broad domain integration confers a distinct advantage, substantially enhancing the predictive capabilities of the student models, each focused on a specific domain.

\begin{table}[]
    \footnotesize
    \renewcommand\arraystretch{1.}
    \setlength{\tabcolsep}{6pt}
    \centering
    \caption{Experiments for demonstrating the cross-model knowledge distillation across domains on EMBARC dataset for Depression Grading, note that both fMRI and EEG are adopted.}
    \begin{tabular}{c|c|c|cc}
        \toprule
        Dataset        &Pre-trained Teacher    &Fine-tuned Student           & AUROC    & Acc         \\
        \midrule
        \multirow{3}{*}{EMBARC}  &\multirow{3}{*}{STF (3 Domains)}   &Spatial (S)      &2\%$\uparrow$    &3\%$\uparrow$      \\
        &   &Temporal (T)      &3\%$\uparrow$    &3\%$\uparrow$  \\
		&   &Frequency (F)     &2\%$\uparrow$    &4\%$\uparrow$  \\
        \bottomrule
    \end{tabular}
    \label{tb_dis}
\end{table}

\subsection{Comparison among the parameters of different models}

\begin{figure}[htbp]
	\centering
	\includegraphics[width=11cm]{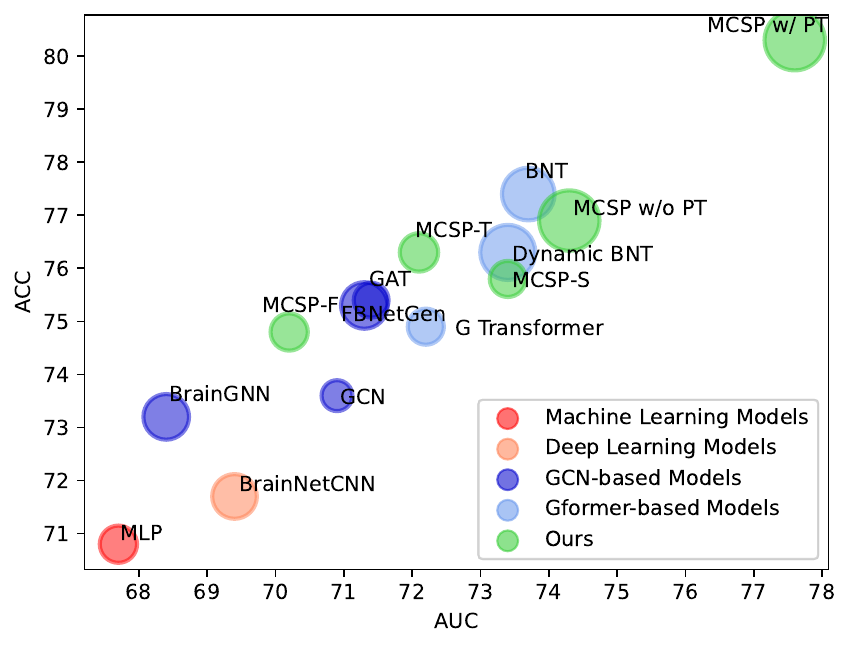}
	\caption{Illustration of Comparison on model parameters and sizes. '-S', '-T' and '-F' refer to the sub-models of our model in three different domains. '-PT' means pre-training. The diameter of each circle represents the number of parameters in the model.}
	\label{exp_param}
\end{figure}

To validate the efficiency of our model, we compared the performance and parameter counts of different models. As shown in Figure \ref{exp_param}, we compared our model's performance and parameter counts with various methods. The results show that although our model is composed of three smaller sub-models and has a slightly higher parameter count than previous methods, its performance significantly surpasses that of other models, especially after undergoing large-scale pre-training. As illustrated in the figure, our three sub-models, each tailored for a specific domain, achieved performance comparable to state-of-the-art methods within their respective domains. However, by integrating these models through a cross-domain self-supervised loss during pre-training, we attained significantly improved results compared to previous methods, despite the increased total number of parameters. Overall, our model consists of a total of 3.4M parameters. Specifically, the sub-model for the spatial domain contains 1.2M parameters, the sub-model for the temporal domain has 1.1M parameters, and the sub-model for the frequency domain includes 1.1M parameters.

\section{Conclusion}
\label{sec:conclusion}

In our study, we developed a Multi-modal Cross-domain Self-supervised Pre-training model designed to integrate multimodal and multidomain data efficiently. We adapted the principles of pretraining to the realm of neuroimaging analysis, allowing for universal pretraining across diverse modalities and domains. Our approach incorporated domain-specific data augmentation to enhance feature representation within each domains. We then harnessed a cross-domain contrastive learning loss function to foster the enhancement of feature similarity, both within and across these domains and modalities. Our rigorous evaluations demonstrated the model's exceptional adaptability across various scenarios, adeptly unlocking the complexities of neural datasets. This universality earmarks our model as a robust tool for a wide array of neuroimaging endeavors, enabling advanced investigations into the multifaceted dynamics of brain functionality and pathology.

\section{Acknowledgment}
This work was supported by NIH grant nos. R01MH129694, R21MH130956, R21AG080425, Alzheimer’s Association Grant (AARG-22-972541), and Lehigh University FIG (FIGAWD35), CORE, and Accelerator grants. Portions of this research were conducted on Lehigh University’s Research Computing infrastructure partially supported by NSF Award 2019035. G.A.F. was also supported by philanthropic funding and NIH grant nos. R01MH132784 and R01MH125886, and grants from the One Mind - Baszucki Brain Research Fund, the SEAL Future Foundation, and the Brain and Behavior Research Foundation.

\bibliographystyle{unsrt}
\bibliography{ref}

\end{document}